\documentclass[aip,jcp,reprint,floatfix,superscriptaddress,longbibliography,raggedbottom]{revtex4-2}

\usepackage{lmodern}
\usepackage{amsmath,amssymb,bm,graphicx,booktabs,microtype,seqsplit}
\usepackage{xurl}
\usepackage[hidelinks]{hyperref}

\usepackage[english]{babel}
\makeatletter
\let\l@en\l@english

\renewcommand{\frontmatter@thefootnote}{*}
\renewcommand{\frontmatter@makefnmark}{\@textsuperscript{\normalfont\@thefnmark}}
\makeatother
\newcommand{\vect}[1]{\bm{#1}}
\newcommand{\qm}{\mathrm{QM}}
\newcommand{\spring}{\mathrm{spring}}

\newcommand{\angstrom}{\text{\AA}}

\makeatletter
\def\section{\@startsection{section}{1}{\z@}
 {12pt plus 2pt minus 2pt}{6pt}
 {\normalfont\small\sffamily\bfseries\raggedright}}
\def\subsection{\@startsection{subsection}{2}{\z@}
 {10pt plus 2pt minus 2pt}{6pt}
 {\normalfont\small\sffamily\bfseries\raggedright}}
\makeatother

\begin{document}
\title{Low-variance estimators for nuclear quantum mean forces}
\author{Venkat Kapil}
\email{v.kapil@ucl.ac.uk}
\affiliation{Department of Physics and Astronomy, University College London, 7--19 Gordon St, London WC1H 0AH, UK}
\affiliation{Thomas Young Centre and London Centre for Nanotechnology, 9 Gordon St, London WC1H 0AH, UK}

\begin{abstract}
Quantum nuclear effects on the statistical properties of position-dependent operators can be rigorously captured by molecular dynamics on the learnt potential of mean force of a single path-integral replica.
However, the primitive mean-force averages obtained from path integral molecular dynamics with that replica held fixed are noisy.
We derive a ``scaled gradient" estimator with zero variance for a one-dimensional harmonic oscillator. 
On our molecular benchmark systems,  the Zundel cation at 100\,K and 300\,K and bulk water at 300\,K we observe nearly 5 to 50 times reduction in sampling time.
The resulting potentials can be trained on these forces like standard machine learning interatomic potentials and yield stable production simulations from which quantum expectation values of position-dependent observables are obtained as classical ensemble averages.
\end{abstract}

\maketitle

\begin{figure*}[!tp]
 \includegraphics[width=\textwidth]{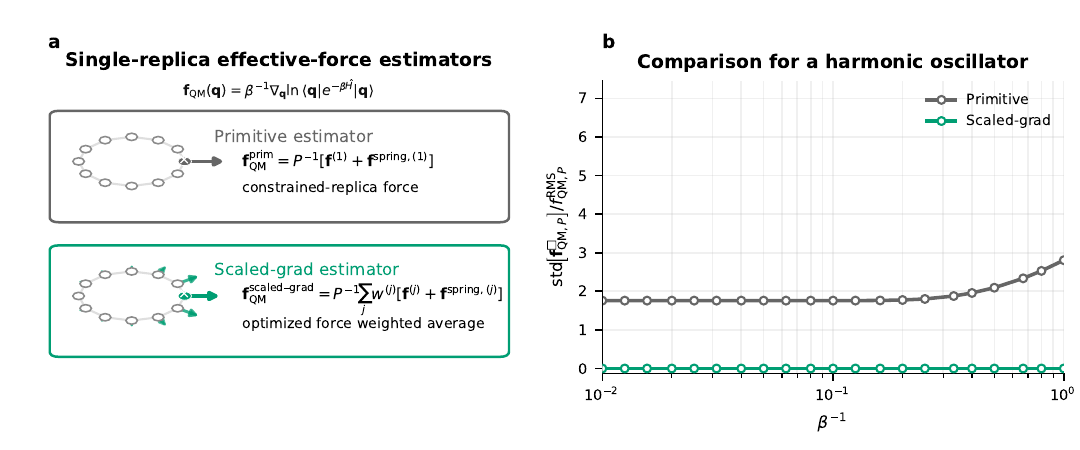}
 \caption{(a) A schematic for the primitive and the scaled gradient estimators. The cross marks the constrained replica while $\vect f$ and $\vect f+\vect f^{\spring}$ denote physical and total ring-polymer forces. Weights are fixed while each estimator is evaluated; (b) Standard deviation of the mean force divided by the root-mean-square force, for a harmonic oscillator with $m=\hbar=\omega=1$, evaluated at selected integer $P$ setting $\beta^{-1}=4/P$ for convergence. Lines are guides to the eye.}
 \label{fig:estimators}
\end{figure*}

\section{Introduction}

Path integral molecular dynamics (PIMD) yields systematically converged  quantum Boltzmann statistics, however, at the overhead of simulating many classical replicas of a system.\cite{feynman_quantum_1965, chandler_exploiting_1981, parrinello_study_1984} 
Existing acceleration strategies address several distinct sources of computational cost.  
Higher-order factorizations of the Boltzmann operator~\cite{takahashi_monte_1984, suzuki_hybrid_1995, chin_symplectic_1997, jang_applications_2001, perez_improving_2011, kapil_high_2016, kapil_modeling_2019}, truncated cumulant expansions of the Boltzmann trace~\cite{poltavsky_accurate_2020, poltavsky_modeling_2016} and non-equilibrium colored-noise thermostats that mimic quantum fluctuations~\cite{ceriotti_nuclear_2009, dammak_quantum_2009, ceriotti_accelerating_2011, ceriotti_efficient_2012,  brieuc_quantum_2016} reduce the number of replicas in the PIMD Hamiltonian.
Approaches such as ring polymer contraction~\cite{markland_efficient_2008} and multiple-time-stepping~\cite{tuckerman_reversible_1992, kapil_accurate_2016, marsalek_quantum_2017} reduce how frequently total energy and gradients are evaluated.

An alternate acceleration strategy is coarse-graining i.e. to integrate out  the replicas of the system that are strictly not necessary for estimating observables and evaluating the resulting effective potential~\cite{cao_new_1993, sinitskiy_reductionist_2015, ryu_coarse-graining_2022, loose_centroid_2022, musil_quantum_2022, zaporozhets_accurate_2024, shen_rigorous_2026}.
Integrating out all but the centroid of the replicas underlies centroid molecular dynamics~\cite{cao_formulation_1994}, which approximates dynamical properties.
In what follows, we will refer to this approach as ``centroid" coarse-graining.   
Integrating out all but one arbitrary replica has been shown to provide a rigorous route to equilibrium structural properties~\cite{zaporozhets_accurate_2024} and equivalent to coarse graining with respect to the starting point of imaginary-time paths~\cite{sadhasivam_effective_2026}. 
In what follows, we will refer to this approach as ``starting point" coarse-graining.

Centroid coarse-graining has been shown to be tractable using bottom-up force-matching.
With the effective potential represented by an empirical force field, the approach was referred to as fast centroid molecular dynamics~\cite{hone_fast_2005, fletcher_fast_2021}.
More recently, with the effective potential represented a machine learning interatomic potential (MLIP), the approach has been referred to as path-integral coarse-graining simulations (PIGS)~\cite{loose_centroid_2022, musil_quantum_2022, zaporozhets_accurate_2024, shen_rigorous_2026}.
Centroid coarse-graining has proven successful as its potential of mean force is often a small correction to the physical potential, with force corrections vanishing at high temperatures and for stiff harmonic modes at all temperatures.~\cite{feynman_quantum_1965}
Moreover, the primitive estimator of the centroid mean force has a low variance.

The same is not true for starting point coarse-graining.
The primitive estimator calculating the mean force contains a spring-force contribution that is absent from the centroid estimator~\cite{zaporozhets_accurate_2024}, and its variance can therefore be much higher with an explicit temperature dependence.
Moreover, at high temperatures, the corresponding potential of mean force approaches the physical potential more slowly than its centroid counterpart and remains significantly different from the physical potential even in the harmonic regime~\cite{sadhasivam_effective_2026}.
The difficulty of learning this effective potential using the PIGS approach was noted in Ref.~\citenum{zaporozhets_accurate_2024}. 
They demonstrated successful generalization for a water molecule, while the larger Zundel system required more training data and simulations of bulk water observed stability issues.
These difficulties also motivate alternative routes to quantum statistical sampling, such as generative Gibbs sampling~\cite{wang_quantum_2026} which can be trained using classical molecular dynamics data.

In this work, we derive a low-variance ``scaled gradient" estimator for the mean forces acting on a replica. 
This estimator yields zero variance for a one-dimensional harmonic oscillator and reduces the estimated constrained-sampling time by factors $\approx$ 5 to 50 for the systems reported here.
The reduced variance of the effective-force estimators enables the training of machine-learned effective potentials that yield substantially more stable production simulations than previous approaches based on the primitive estimator. 
This opens a promising route towards accurate and scalable simulations incorporating quantum nuclear effects using classical molecular dynamics.
\section{Theory and methods}

\subsection{Fixed-replica constrained ensemble}

The expression for the effective potential governing quantum statistics for a Hamiltonian $\hat{H}$ is
\begin{equation}
U_{\textrm{QM}}(\vect q; \beta)
=-\beta^{-1}\log
{\langle \vect q|e^{-\beta\hat H}|\vect q\rangle}
+ \beta^{-1}\log {Z(\beta)}
\label{eq:ueff}
\end{equation}
where $\mathbf{q}$ is the position vector of the system, $\beta$ is the inverse temperature, and $Z(\beta)$ is the canonical quantum partition function.
The corresponding imaginary-time discretized path-integral expression is
\begin{equation}
\begin{aligned}
&U_{\textrm{QM},P}(\mathbf q;\beta)
=-\beta^{-1}\log\Bigg[
\int\prod_{j=1}^{P}\text{d}\mathbf q^{(j)}
\\
&\qquad
\delta^{(3N)}\left(\mathbf q-\mathbf q^{(1)}\right)
e^{-\beta_P U_P\left(\{\mathbf q^{(j)}\}\right)}
\Bigg]+\mathrm{const.}
\end{aligned}
\label{eq:Ueff_pi}
\end{equation}
where $P$ is the number of replicas, $\beta_P \equiv \beta/P$, $\mathbf q^{(j)}$ is the position vector of the $j$-th replica, and we choose to constrain the arbitrarily selected first replica at $\mathbf{q}$.
The potential term 
\[
U_P\left(\{\mathbf q^{(j)}\}\right) =
\sum_{j=1}^{P}
\left[
U(\mathbf q^{(j)}) 
+
\sum_{i=1}^{N}
\frac{m_i |\mathbf q_i^{(j)}-\mathbf q_i^{(j+1)}|^2}{2\beta_P^2\hbar^2}
\right].
\]
includes the sum of the physical potentials on all replicas and a spring term. 
Implementation of Eq.~\ref{eq:Ueff_pi} requires PIMD with the first replica constrained at $\vect q$. 

\subsection{Primitive estimator for the mean force}

The mean force acting on the first replica can be evaluated using the primitive estimator
\begin{equation}
\vect f_{\qm,P}(\vect q)=P^{-1}\left\langle
\vect f^{(1)} + \vect f^{\spring,(1)}\right\rangle_{\vect q}
\label{eq:constraint}
\end{equation}
where $\vect f^{(1)}$ and $\vect f^{\spring,(1)}$ are respectively the physical and spring forces acting on the first replica, and $\left<\cdot\right>_{\mathbf{q}}$ denotes the canonical ensemble with the first replica constrained at $\vect q$.
A schematic representation of the primitive estimator is shown in Fig.~\ref{fig:estimators}(a).
Unfortunately, the primitive estimator exhibits poor statistical behaviour for certain systems and temperatures because its variance depends strongly on temperature through the spring frequency $\omega_P=P/(\beta\hbar)$.

As shown in Fig.~\ref{fig:estimators}(b) and derived in Appendix~\ref{app:primitive-noise}, we quantify this behaviour for a one-dimensional harmonic oscillator with unit mass and unit frequency using the noise-to-signal ratio of the force estimator. 
This ratio is defined as the standard deviation of the estimator divided by the root-mean-square mean force and for a harmonic oscillator it is independent of the replica's absolute position.
For the convergence rule used in Fig.~\ref{fig:estimators}(b), $P=4\beta\hbar\omega$, this ratio approaches 1.76 as $\beta\hbar\omega\to\infty$ and increases as $\beta\hbar\omega$ decreases over the range shown.
The primitive estimator can therefore require long constrained PIMD simulations for systems containing both high- and low-frequency modes, due to the poor statistical properties of the effective forces associated with the low-frequency modes.
Since only a fraction of the modes in a typical system are strongly quantized, much of the computational cost of the primitive estimator arises from modes for which nuclear quantum effects are small. 

\subsection{Scaled gradient estimator for the mean force}

A more efficient estimator can be derived by noting that the average total force on each unconstrained replica vanishes in the ensemble defined in Eq.~\ref{eq:constraint} (see Appendix~\ref{app:zero-mean}).
Consequently, the estimator
\begin{equation}
 \vect f_{\qm,P}^{\mathrm{scaled~gradient}}
 =
 P^{-1}\sum_{j=1}^{P} w^{(j)}
 \left[
 \vect f^{(j)}+\vect f^{\spring,(j)}
 \right]
 \label{eq:sg_estimator}
\end{equation}
has the same expectation value as the primitive estimator for any set of fixed weights satisfying $w^{(1)}=1$.
The remaining $P-1$ weights can therefore be chosen to minimize the variance of the estimator,
\begin{equation}
 \mathbf w_{\mathrm{opt}}
 =
 \underset{\mathbf w:\,w^{(1)}=1}{\operatorname{argmin}}
 \operatorname{Var}
 \left[
\vect f_{\qm,P}^{\mathrm{scaled~gradient}}
 \right].
 \label{eq:optimal_weights}
\end{equation}
Here $\operatorname{Var}$ denotes the mean variance over $3N$ Cartesian force components.
We refer to Eq.~\ref{eq:sg_estimator} as the ``scaled gradient" estimator, following a related estimator based on the same idea that we previously developed for open paths~\cite{kapil_anisotropy_2018}, and present its schematic diagram in Fig.~\ref{fig:estimators}(a).

For a one-dimensional harmonic oscillator, the weights that minimize the variance can be obtained analytically, yielding a zero-variance estimator, as shown in Appendix~\ref{app:harmonic}.
For a harmonic oscillator with mass $m$ and frequency $\omega$, the optimal weights are
\begin{equation}
w_{\mathrm{opt}}^{(j)}
=
\frac{
\cosh\left[\left(P/2-j+1\right)\kappa\right]
}{
\cosh\left(P\kappa/2\right)
},
\label{eq:ho_optimal_weights}
\end{equation}
where $\kappa=\operatorname{arcosh}\left[1+\frac{1}{2}\left(\beta_P\hbar\omega\right)^2\right]$.
Fig.~\ref{fig:estimators}(b) compares the noise-to-signal ratio of this estimator with that of the primitive estimator for the harmonic oscillator.
For realistic systems and finite simulation lengths, however, the optimal weights must be estimated from sampled force correlations, and the achievable variance reduction therefore depends on the quality of configurational sampling.

\subsection{PIGS + the scaled gradient force estimator}

For any system for which an analytical treatment is not possible, the mean QM force for a given configuration $\mathbf{q}$ must be estimated from constrained PIMD simulations with one replica constrained at $\mathbf{q}$.
The total forces on all replicas are sampled along the MD trajectory and averaged over non-overlapping trajectory blocks.
For block $r$, we write this average as
\begin{equation}
\vect F_r^{(j)}=
\left\langle\vect f^{(j)}+\vect f^{\spring,(j)}\right\rangle_r,
\qquad j=1,\ldots,P,
\label{eq:block-total-force}
\end{equation}
where $\langle\cdot\rangle_r$ denotes the time average within that block.
The blocks are divided equally into two sets.
The first set is used to fit the weights and determine the regularization parameter, while the second set is reserved for evaluating the resulting estimator.
To determine the regularization parameter, the fitting set is further divided into two equal training and validation subsets.
The training subset is used to fit $\mathbf w(\lambda)$ by minimizing Eq.~\ref{eq:optimal_weights} with an additional $L_2$ regularization term, which leads to the closed-form solution
\begin{equation}
\mathbf w(\lambda) = -(\mathbf C+\lambda\mathbf I)^{-1}\mathbf b,
\label{eq:ridge-weights}
\end{equation}
where $\lambda$ is a regularization hyperparameter.
For any set $\mathcal B$ of fitting blocks, let $\overline{\vect F}_{\mathcal B}^{(j)}=|\mathcal B|^{-1}\sum_{r\in\mathcal B}\vect F_r^{(j)}$.
The covariance matrix $\mathbf C$ of the unconstrained-replica forces and their covariance vector $\mathbf b$ with the constrained-replica force are
\begin{align}
C_{ij}
&=\frac{\displaystyle\sum_{r\in\mathcal B}
(\vect F_r^{(i)}-\overline{\vect F}_{\mathcal B}^{(i)})\cdot
(\vect F_r^{(j)}-\overline{\vect F}_{\mathcal B}^{(j)})}
{3N(|\mathcal B|-1)},\nonumber\\
b_i
&=\frac{\displaystyle\sum_{r\in\mathcal B}
(\vect F_r^{(i)}-\overline{\vect F}_{\mathcal B}^{(i)})\cdot
(\vect F_r^{(1)}-\overline{\vect F}_{\mathcal B}^{(1)})}
{3N(|\mathcal B|-1)}.
\label{eq:block-covariances}
\end{align}
Here $i,j=2,\ldots,P$.
Candidate values of $\lambda$ are evaluated by applying the corresponding weights to the validation blocks and computing the variance across those blocks, as defined in Appendix~\ref{app:scaled-grad-weights}.
The roles of the training and validation subsets are then exchanged and the procedure is repeated.
The value of $\lambda$ is selected from the combined validation results.
Using the selected value of $\lambda$, $\mathbf C$ and $\mathbf b$ are recomputed using all blocks in the fitting set and the resulting weights are applied to the second, previously unused set of blocks.
Finally, the two original sets are exchanged: the second set is used to select $\lambda$ and fit the weights following the same procedure, and the resulting weights are applied to the first set.
This procedure yields one force estimate for every trajectory block using weights determined without using that block either to select the regularization parameter or to fit the weights.

The procedure described above is applied independently to a set of configurations $\{\mathbf q\}$ to yield a set of quantum mean forces $\{\vect f_{\qm,P}(\mathbf q)\}$.
The resulting configuration-force pairs form the dataset used to learn the effective potential at a given inverse temperature.
The dataset is divided into training and validation sets, and a conservative MACE potential~\cite{batatia_mace_2022} $U_\theta$ is fitted to the training configurations using a forces loss
\begin{equation}
\mathcal{L}(\theta)
=
\frac{1}{N_{\mathrm{train}}}
\sum_{\mathbf q \in \mathcal{D}_{\mathrm{train}}}
\left|
-\nabla_{\mathbf q} U_\theta(\mathbf q)
-
\vect f_{\qm,P}(\mathbf q)
\right|^2.
\end{equation}
The model that minimizes the corresponding force loss on the validation set is selected for production simulations.
We learn the effective potential directly from the quantum mean-force labels and do not use a differential-learning scheme with respect to the underlying physical potential energy surface.
Classical sampling of the resulting $U_\theta$ then approximates the target single-replica quantum distribution.
We will refer to these simulations as PIGS.

\section{Results}

\subsection{Zundel cation}

We first  implement a replica-level constraint in the i-PI code~\cite{litman_i-pi_2024}. 
All PIMD simulations for calculating the quantum mean force and all standard molecular dynamics and PIMD simulations were performed with the i-PI code.  

We start by considering the case of the Zundel cation at 100 and 300\,K which exhibits proton delocalization between two oxygen atoms. 
Initial configurations are sampled from a 100 ps long unconstrained PIMD simulation at 300 K. 
For 64 configurations randomly sampled from the replica trajectories, bead-constrained PIMD simulations are performed at each temperature for 7\,ps, with 2\,ps equilibration and 5\,ps production. 
Total forces are sampled every 10\,fs, and each production trajectory is split into 20 blocks of 0.25\,ps, which are used to estimate the mean force with the scaled gradient estimator, as outlined earlier, and with the primitive estimator as a control. 
The resulting dataset for each temperature comprises 48 training, eight validation, and eight test configurations.

\begin{figure}[!tp]
 \includegraphics[width=\columnwidth]{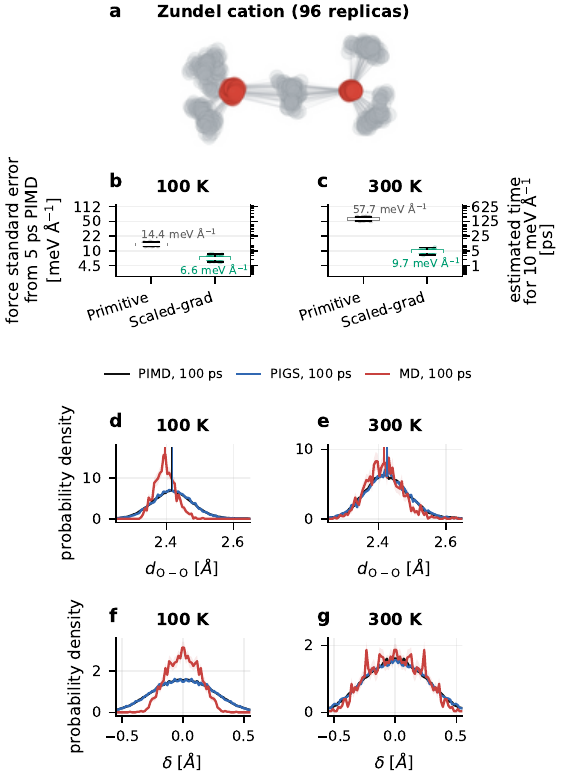}
 \caption{Panel (a) overlays all 96 replicas of a representative 100 K PIMD configuration. Panels (b) and (c) compare the standard error for the  mean forces associated with the primitive and scaled gradient estimators at 100 and 300 K, respectively; boxes show the interquartile range, white lines show the median, and points show individual configurations. The logarithmic right axes give the estimated thermostatted time for which a constrained path integral molecular dynamics needs to be run for a 10 meV~$\angstrom^{-1}$ force error estimated from a $t^{-1/2}$ dependence of the standard error.
 Panels (d) and (e) show the O--O distributions at 100 and 300 K, respectively, while panels (f) and (g) show the distributions associated with the proton transfer coordinate. Black, blue, and red denote PIMD, PIGS, and classical MD, respectively. All structural distributions use the full $0\leq t<100$ ps interval; shading shows the standard error from fifty 2 ps blocks. 
 }
 \label{fig:zundel}
\end{figure}

As shown in Fig.~\ref{fig:zundel}(b) and (c), the scaled gradient estimator substantially reduces the standard error associated with the quantum mean force.
The median standard error decreases from 14.4 to 6.6 meV~$\angstrom^{-1}$ at 100 K and from 57.7 to 9.7 meV~$\angstrom^{-1}$ at 300 K.
To quantify what this improvement in statistical efficiency corresponds to in terms of PIMD simulation time, we consider a target force uncertainty of $\approx 10$ meV~$\angstrom^{-1}$, comparable to the typical force error of MLIPs trained on density-functional-theory data, and estimate the required simulation time $t$ assuming a $t^{-1/2}$ convergence of the standard error.
The corresponding median sampling times required to reach 10 meV~$\angstrom^{-1}$ are 10.3 and 2.2 ps at 100 K, and 166.3 and 4.7 ps at 300 K, for the primitive and scaled gradient estimators, respectively.
This corresponds to reductions in the required sampling time by factors of approximately 4.7 and 35 at 100 and 300 K, respectively.
The absolute computational savings at 100 K are further amplified by the higher cost per PIMD step, as these simulations require three times as many replicas as those at 300 K.

\begin{figure*}[!tp]
 \includegraphics[width=\textwidth]{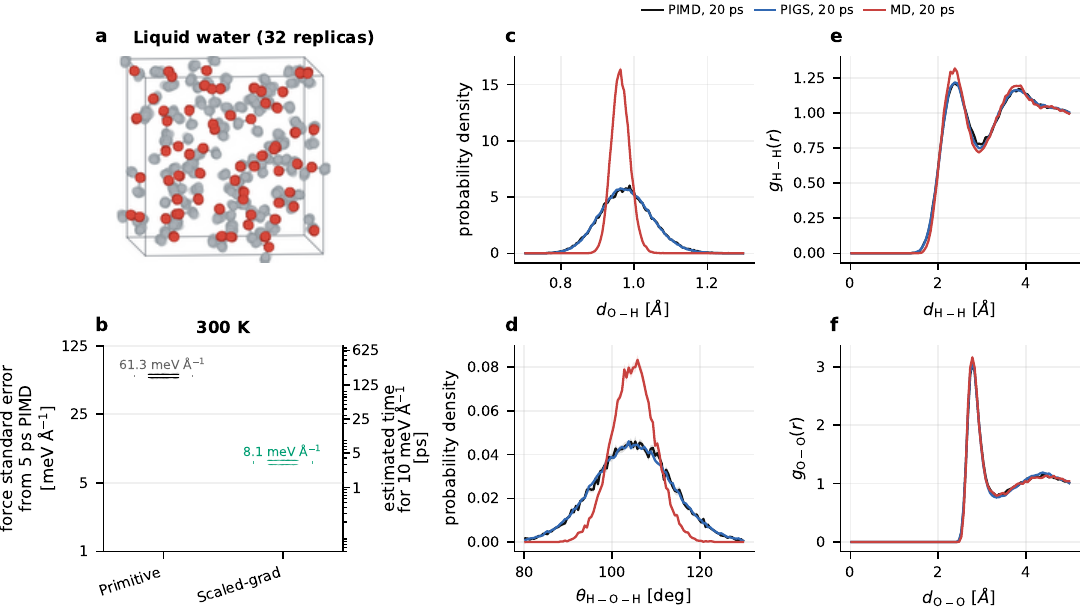}
 \caption{Panel (a) overlays all 32 replicas of a representative configuration from the 64-molecule PIMD reference. Panel (b) compares the standard error for the mean forces associated with the primitive and scaled gradient estimators; boxes show the interquartile range, white lines show the median, and points show individual configurations. The logarithmic right axes give the estimated thermostatted time for which a constrained path integral molecular dynamics needs to be run for a 10 meV~$\angstrom^{-1}$ force error estimated from a $t^{-1/2}$ dependence of the standard error. Panels (c) and (d) show the intramolecular O--H distance and $\theta_{\mathrm{H-O-H}}$, while panels (e) and (f) show the intermolecular H--H and O--O radial distributions. Black, blue, and red denote 32-replica PIMD using replica 0, PIGS using the ridge-trained epoch-80 model with 128 configurations, and classical q-TIP4P/F MD, respectively. All curves use the full $0\leq t<20$ ps interval. Shading shows the standard error of the histograms estimated from twenty 1 ps blocks.}
 \label{fig:water}
\end{figure*}

For validation on equilibrium properties, independent 100 ps PIMD trajectories at 100 and 300 K serve as references.
Within the PIGS approach, the MLIPs fitted to the scaled gradient mean forces are used to perform standard MD simulations of the same duration as the reference PIMD simulations.
We compare two structural distributions: the histogram of the O--O distance, $d_{\textrm{O}-\textrm{O}}$, and the histogram of the proton-transfer coordinate, $\delta$.
We also perform classical MD simulations of the same duration as the reference PIMD simulations to quantify quantum nuclear effects.
As shown in Fig.~\ref{fig:zundel}(d) and (f), at 100\,K both distributions are broadened in PIMD relative to classical MD, while the PIGS distributions are in quantitative agreement with PIMD. 
In addition, quantum nuclear effects elongate the O--O distance and the PIGS approach is able to quantitatively determine it. 
As shown in Fig.~\ref{fig:zundel}(e) and (g), at 300\,K the differences between the PIMD and classical MD distributions (width and mean values) are smaller than at 100\,K; nevertheless, the PIGS distributions remain in quantitative agreement with PIMD.

\subsection{Liquid water}

We next consider liquid water described by the q-TIP4P/F potential~\cite{habershon_competing_2009} at 300 K.
The system contains 64 water molecules in a cubic cell of side 12.4138~\angstrom and is simulated using 32 replicas.
The labeling set contains 128 configurations, of which 112 are used for training, eight for validation, and eight for testing.
Each constrained calculation uses the same 2 ps equilibration, 5 ps production, force-sampling interval, and block construction as the Zundel calculations.
As shown in Fig.~\ref{fig:water}(b), the scaled gradient estimator reduces the median RMS standard error associated with the quantum mean force from 61.3 to 8.1 meV~$\angstrom^{-1}$.
Assuming $t^{-1/2}$ convergence of the standard error, the corresponding median sampling time required to reach 10 meV~$\angstrom^{-1}$ decreases from 188.1 to 3.3 ps.
This corresponds to a reduction in the required sampling time by a factor of approximately 57.5.

Within the PIGS approach, the MLIP fitted to the scaled gradient mean forces is used to perform a standard MD simulation, and its structural distributions are compared with the PIMD reference.
As shown in Fig.~\ref{fig:water}(c) and (d), PIGS reproduces the broad intramolecular O--H and $\theta_{\mathrm{H-O-H}}$ distributions, while classical MD produces substantially narrower distributions.
The PIGS intermolecular H--H radial distribution is closer to PIMD than classical MD, whereas its O--O discrepancy is comparable to that of classical MD, as shown in Fig.~\ref{fig:water}(e) and (f).
Finally, the model also produces a stable 100\,ps trajectory, circumventing stability issues as noted in Ref.~\citenum{zaporozhets_accurate_2024}.

\section{Conclusions}
Starting-point coarse graining offers a way to obtain quantum equilibrium structure without carrying a full ring polymer through every production step, but the primitive mean-force estimator makes the required force labels expensive to obtain.
We have shown that the zero-mean total forces on the unconstrained replicas can be combined to reduce this statistical noise.
The scaled gradient estimator has zero sampling variance for a one-dimensional harmonic oscillator and reduces the estimated constrained-labeling time by nearly 5 to 50 times in the Zundel and liquid-water tests.
The estimator uses forces already available in a constrained PIMD simulation and does not depend on a particular physical potential or learning architecture, making it simple and broadly applicable.
It provides a route to building larger datasets of quantum mean forces at lower labeling cost and reusing the learned potential in long or large-scale PIGS production simulations.
If the learned potential remains accurate over the configurations encountered in production, the computational cost of PIMD labeling can be amortized across these simulations.
Just as foundation MLIPs~\cite{creed_six_2026, yuan_foundation_2026} have extended first-principles-based simulations to larger systems and longer times, we believe our work could be a step towards a foundation PIGS models for equilibrium nuclear quantum effects accessible at those scales at the same computational cost.

\begin{acknowledgments}
The author acknowledges support from University College London startup funds.
\end{acknowledgments}

\section*{Data Availability Statement}
The datasets and trained models supporting the findings of this work are publicly available at \url{https://github.com/venkatkapil24/data-low-variance-starting-point-pigs}.

\section*{Code Availability Statement}
A \href{https://github.com/i-pi/i-pi/pull/592}{pull request (\#592)} adding the replica constraints required for these calculations has been opened in i-PI, and we expect it to be merged.
The scripts for calculating the scaled gradient mean forces and reproducing the workflow are available in the repository linked above.

\section*{Contributions and AI Use}
The author conceived the scaled gradient estimator and directed the research. GPT-5.6-Sol-High + Codex harness, carried out the derivations, implemented the method in i-PI, ran the simulations, wrote the analysis scripts, generated the figures, and prepared the first draft of the manuscript. The author manually verified the derivations, revised the manuscript, and takes responsibility for the final work.

%

\onecolumngrid
\appendix
\setcounter{secnumdepth}{3}

\section{Primitive estimator: harmonic oscillator analytic exercise}
\label{app:primitive-noise}

In this section we derive the mean force and its root mean square, followed by the variance of the primitive force estimator (Eq.~~\ref{eq:constraint}) for a one-dimensional harmonic oscillator in the replica-constrained ensemble.
We then compare the estimator's standard deviation with the root-mean-square mean force.

\subsection{The mean force}

\subsubsection{Definition}

For an oscillator of mass $m$ and frequency $\omega$, the Hamiltonian is
$\hat H=\hat p^2/(2m)+m\omega^2\hat q^2/2$.
Substituting its physical potential into the ring-polymer potential gives
\begin{equation}
U_P=\sum_{j=1}^{P}\left[
m\omega^2(q^{(j)})^2/2
+m\omega_P^2(q^{(j)}-q^{(j+1)})^2/2
\right],
\end{equation}
with the constraint $q^{(1)}=q$.
The primitive force estimator is
\begin{equation}
f_{\qm,P}^{\mathrm{prim}}(q,\{q^{(j)}\})
=-m~~\left[
\omega^2q+\omega_P^2(2q-q^{(2)}-q^{(P)})
\right]/P.
\label{eq:s-harmonic-primitive-sample}
\end{equation}

The mean force is the ensemble average of this estimator:
\begin{equation}
f_{\qm,P}(q)=
\frac{
\displaystyle\int\prod_{j=2}^{P}\textrm{d}q^{(j)}~~
-m\left[
\omega^2q+\omega_P^2(2q-q^{(2)}-q^{(P)})
\right]/P~~
e^{-\beta_P\sum_{j=1}^{P}
\left[
m\omega^2(q^{(j)})^2/2
+m\omega_P^2(q^{(j)}-q^{(j+1)})^2/2
\right]}
}{
\displaystyle\int\prod_{j=2}^{P}\textrm{d}q^{(j)}~~
e^{-\beta_P\sum_{j=1}^{P}
\left[
m\omega^2(q^{(j)})^2/2
+m\omega_P^2(q^{(j)}-q^{(j+1)})^2/2
\right]}
}.
\label{eq:s-harmonic-force-average}
\end{equation}

\subsubsection{Change of variables}

We rewrite the ring-polymer potential as
\begin{equation}
U_P=\frac{m}{2}\left[
\vect x^T\vect H\vect x
-2\omega_P^2q~~\vect e^T\vect x
+(\omega^2+2\omega_P^2)q^2
\right],
\label{eq:s-harmonic-action}
\end{equation}
where $\vect x\equiv(q^{(2)},\ldots,q^{(P)})^{\textrm{T}}$ is the position vector associated with the unconstrained replicas and
\begin{align}
H_{ij}
&=(\omega^2+2\omega_P^2)\delta_{ij}
-\omega_P^2(\delta_{i,j+1}+\delta_{i,j-1}),\nonumber\\
e_i
&=\delta_{i1}+\delta_{i,P-1},
\qquad i,j=1,\ldots,P-1.
\label{eq:s-harmonic-matrix}
\end{align}
We shift the integration variable (with a transformation with a unit Jacobian) to complete the square and rewrite the potential term and primitive force estimator terms
\[
\vect y=\vect x-\omega_P^2\vect H^{-1}\vect e~~q,
\]
\[
U_P=\frac{m}{2}\left[
\vect y^T\vect H\vect y
+\left(\omega^2+2\omega_P^2
-\omega_P^4\vect e^T\vect H^{-1}\vect e\right)q^2
\right].
\]
\[
f_{\qm,P}^{\mathrm{prim}}(q,\vect y)
=-\frac{m}{P}\left[
\left(\omega^2+2\omega_P^2
-\omega_P^4\vect e^T\vect H^{-1}\vect e\right)q
-\omega_P^2\vect e^T\vect y
\right].
\]
Its ensemble average of the primitive force estimator is
\begin{equation}
f_{\qm,P}(q)=
-\frac{m}{P}\left[
\left(\omega^2+2\omega_P^2
-\omega_P^4\vect e^T\vect H^{-1}\vect e\right)q
\right]
+
\frac{
\displaystyle\int \textrm{d}\vect y~~
m\omega_P^2\vect e^T\vect y/P~~
e^{-\beta_P(m/2)\left[
\vect y^T\vect H\vect y+
(\omega^2+2\omega_P^2-\omega_P^4\vect e^T\vect H^{-1}\vect e)q^2
\right]}
}{
\displaystyle\int \textrm{d}\vect y~~
e^{-\beta_P(m/2)\left[
\vect y^T\vect H\vect y+
(\omega^2+2\omega_P^2-\omega_P^4\vect e^T\vect H^{-1}\vect e)q^2
\right]}
}.
\label{eq:s-harmonic-force-average-shifted}
\end{equation}

\subsubsection{Evaluation of the Gaussian integrals}

The denominator of Eq.~~\eqref{eq:s-harmonic-force-average-shifted} evaluates to
\begin{equation}
[2\pi/(\beta_P m)]^{(P-1)/2}
(\det\vect H)^{-1/2}
e^{-\beta_P(m/2)
(\omega^2+2\omega_P^2-\omega_P^4\vect e^T\vect H^{-1}\vect e)q^2}.
\label{eq:s-harmonic-normalization}
\end{equation}
Its explicit value is unnecessary for the mean-force result as the term in the primitive estimator proportional to $\vect e^T\vect y$ is odd in $\vect y$ and integrates to zero.
We use the denominator below to calculate the root-mean-square mean force.
The mean force is therefore
\begin{equation}
f_{\qm,P}(q)
=-\frac{m}{P}
\left(\omega^2+2\omega_P^2
-\omega_P^4\vect e^T\vect H^{-1}\vect e\right)q.
\label{eq:s-harmonic-effective}
\end{equation}

\subsection{The root-mean-square mean force}

\subsubsection{Definition}

The root-mean-square mean force averages the square of
$f_{\qm,P}(q)$ over the position $q$ of the first replica is 
\begin{equation}
f_{\qm,P}^{\mathrm{RMS}}
=
\left[
\frac{
\displaystyle\int\textrm{d}q~~[f_{\qm,P}(q)]^2
\int\prod_{j=2}^{P}\textrm{d}q^{(j)}~~
e^{-\beta_P\sum_{j=1}^{P}
\left[
m\omega^2(q^{(j)})^2/2
+m\omega_P^2(q^{(j)}-q^{(j+1)})^2/2
\right]}
}{
\displaystyle\int\textrm{d}q
\int\prod_{j=2}^{P}\textrm{d}q^{(j)}~~
e^{-\beta_P\sum_{j=1}^{P}
\left[
m\omega^2(q^{(j)})^2/2
+m\omega_P^2(q^{(j)}-q^{(j+1)})^2/2
\right]}
}
\right]^{1/2}.
\label{eq:s-harmonic-rms-definition}
\end{equation}

\subsubsection{Change of variables}

At each value of $q$, we use the same change of variables
$\vect y=\vect x-\omega_P^2\vect H^{-1}\vect e~~q$ as before.
Eq.~~\eqref{eq:s-harmonic-rms-definition} becomes
\begin{equation}
f_{\qm,P}^{\mathrm{RMS}}
=
\left[
\frac{
\displaystyle\int\textrm{d}q~~[f_{\qm,P}(q)]^2
\int\textrm{d}\vect y~~
e^{-\beta_P(m/2)\left[
\vect y^T\vect H\vect y+
(\omega^2+2\omega_P^2-\omega_P^4\vect e^T\vect H^{-1}\vect e)q^2
\right]}
}{
\displaystyle\int\textrm{d}q
\int\textrm{d}\vect y~~
e^{-\beta_P(m/2)\left[
\vect y^T\vect H\vect y+
(\omega^2+2\omega_P^2-\omega_P^4\vect e^T\vect H^{-1}\vect e)q^2
\right]}
}
\right]^{1/2}.
\label{eq:s-harmonic-rms-shifted}
\end{equation}

\subsubsection{Evaluation of the Gaussian integrals}

The integral over $\vect y$ is the same in the numerator and
denominator of Eq.~~\eqref{eq:s-harmonic-rms-shifted}.
Its value is independent of $q$ apart from the exponential factor shown
in Eq.~~\eqref{eq:s-harmonic-normalization}, so the remaining average is
\begin{equation}
\left(f_{\qm,P}^{\mathrm{RMS}}\right)^2
=
\frac{
\displaystyle\int\textrm{d}q~~[f_{\qm,P}(q)]^2
e^{-\beta_P(m/2)
(\omega^2+2\omega_P^2-\omega_P^4\vect e^T\vect H^{-1}\vect e)q^2}
}{
\displaystyle\int\textrm{d}q~~
e^{-\beta_P(m/2)
(\omega^2+2\omega_P^2-\omega_P^4\vect e^T\vect H^{-1}\vect e)q^2}
}.
\label{eq:s-harmonic-rms-position-integral}
\end{equation}
Substituting the mean force from Eq.~~\eqref{eq:s-harmonic-effective}
requires the Gaussian integral
\begin{equation}
\frac{
\displaystyle\int\textrm{d}q~~q^2
e^{-\beta_P(m/2)
(\omega^2+2\omega_P^2-\omega_P^4\vect e^T\vect H^{-1}\vect e)q^2}
}{
\displaystyle\int\textrm{d}q~~
e^{-\beta_P(m/2)
(\omega^2+2\omega_P^2-\omega_P^4\vect e^T\vect H^{-1}\vect e)q^2}
}
=
\frac{1}{
\beta_Pm
(\omega^2+2\omega_P^2-\omega_P^4\vect e^T\vect H^{-1}\vect e)
}.
\label{eq:s-harmonic-position-average}
\end{equation}
Using $\beta_P=\beta/P$, the root-mean-square mean force is therefore
\begin{equation}
f_{\qm,P}^{\mathrm{RMS}}
=
\left[
\frac{m}{\beta P}
\left(
\omega^2+2\omega_P^2
-\omega_P^4\vect e^T\vect H^{-1}\vect e
\right)
\right]^{1/2}.
\label{eq:s-harmonic-force-rms}
\end{equation}

\subsection{The variance of the primitive estimator}

\subsubsection{Definition}

At fixed $q$, the variance of the primitive force estimator is the ensemble average of the
squared difference between the estimator and its ensemble average
\begin{equation}
\left[\sigma_{\qm,P}^{f}(q)\right]^2=
\frac{
\displaystyle\int\prod_{j=2}^{P}\textrm{d}q^{(j)}~~
\left[
f_{\qm,P}^{\mathrm{prim}}(q,\{q^{(j)}\})-f_{\qm,P}(q)
\right]^2
e^{-\beta_P\sum_{j=1}^{P}
\left[
m\omega^2(q^{(j)})^2/2
+m\omega_P^2(q^{(j)}-q^{(j+1)})^2/2
\right]}
}{
\displaystyle\int\prod_{j=2}^{P}\textrm{d}q^{(j)}~~
e^{-\beta_P\sum_{j=1}^{P}
\left[
m\omega^2(q^{(j)})^2/2
+m\omega_P^2(q^{(j)}-q^{(j+1)})^2/2
\right]}
}.
\label{eq:s-harmonic-variance-definition}
\end{equation}

\subsubsection{Change of variables}

At fixed $q$, we use the same change of variables
$\vect y=\vect x-\omega_P^2\vect H^{-1}\vect e~~q$
as before.
Equation~~\eqref{eq:s-harmonic-variance-definition} becomes
\begin{equation}
\left[\sigma_{\qm,P}^{f}(q)\right]^2=
\frac{
\displaystyle\int\textrm{d}\vect y~~
\left[
f_{\qm,P}^{\mathrm{prim}}(q,\vect y)-f_{\qm,P}(q)
\right]^2
e^{-\beta_P(m/2)\left[
\vect y^T\vect H\vect y+
(\omega^2+2\omega_P^2-\omega_P^4\vect e^T\vect H^{-1}\vect e)q^2
\right]}
}{
\displaystyle\int\textrm{d}\vect y~~
e^{-\beta_P(m/2)\left[
\vect y^T\vect H\vect y+
(\omega^2+2\omega_P^2-\omega_P^4\vect e^T\vect H^{-1}\vect e)q^2
\right]}
}.
\label{eq:s-harmonic-variance-shifted}
\end{equation}

\subsubsection{Evaluation of the Gaussian integrals}

Subtracting the mean force in
Eq.~~\eqref{eq:s-harmonic-effective} from the transformed primitive
estimator gives
\begin{equation}
f_{\qm,P}^{\mathrm{prim}}(q,\vect y)-f_{\qm,P}(q)
=
\frac{m\omega_P^2}{P}\vect e^T\vect y.
\label{eq:s-harmonic-centered-primitive}
\end{equation}
The factor depending on $q$ is identical in the numerator and
denominator of Eq.~~\eqref{eq:s-harmonic-variance-shifted} and cancels.
The variance is therefore
\begin{equation}
\left[\sigma_{\qm,P}^{f}(q)\right]^2
=
\frac{m^2\omega_P^4}{P^2}
\frac{
\displaystyle\int\textrm{d}\vect y~~
(\vect e^T\vect y)^2
e^{-\beta_P m\vect y^T\vect H\vect y/2}
}{
\displaystyle\int\textrm{d}\vect y~~
e^{-\beta_P m\vect y^T\vect H\vect y/2}
}.
\label{eq:s-harmonic-variance-integral}
\end{equation}
The remaining integral evaluates to
\begin{equation}
\frac{
\displaystyle\int\textrm{d}\vect y~~
(\vect e^T\vect y)^2
e^{-\beta_P m\vect y^T\vect H\vect y/2}
}{
\displaystyle\int\textrm{d}\vect y~~
e^{-\beta_P m\vect y^T\vect H\vect y/2}
}
=
\frac{1}{\beta_P m}\vect e^T\vect H^{-1}\vect e.
\label{eq:s-harmonic-squared-force-integral}
\end{equation}
Substituting this result into
Eq.~~\eqref{eq:s-harmonic-variance-integral} and using
$\beta_P=\beta/P$ gives
\begin{equation}
\left[\sigma_{\qm,P}^{f}(q)\right]^2
=
\frac{m\omega_P^4}{\beta P}
\vect e^T\vect H^{-1}\vect e.
\label{eq:s-harmonic-variance-matrix}
\end{equation}
Interestingly, the variance is independent of $q$ for the harmonic oscillator.

\subsection{The noise-to-signal ratio}

We define the noise-to-signal ratio as the standard deviation
of the ratio
\begin{equation}
\mathrm{NSR}_{\mathrm{prim},P}(q)
=
\frac{\sigma_{\qm,P}^{f}(q)}{f_{\qm,P}^{\mathrm{RMS}}}.
\label{eq:s-harmonic-nsr-definition}
\end{equation}
Substituting Eqs.~~\eqref{eq:s-harmonic-variance-matrix}
and \eqref{eq:s-harmonic-force-rms} gives
\begin{equation}
\mathrm{NSR}_{\mathrm{prim},P}(q)
=
\left[
\frac{
\omega_P^4\vect e^T\vect H^{-1}\vect e
}{
\omega^2+2\omega_P^2
-\omega_P^4\vect e^T\vect H^{-1}\vect e
}
\right]^{1/2}.
\label{eq:s-harmonic-finite-nsr}
\end{equation}
Unsurprisingly, the ratio is independent of $q$.

\subsection{Evaluation of the remaining matrix element}

To evaluate $\vect e^T\vect H^{-1}\vect e$, let
$\vect z=\vect H^{-1}\vect e$.
The tridiagonal matrix in Eq.~~\eqref{eq:s-harmonic-matrix} makes
$\vect H\vect z=\vect e$ a second-order difference equation.
For $i=1,\ldots,P-1$, it can be written as
\begin{equation}
(\omega^2+2\omega_P^2)z_i
-\omega_P^2z_{i-1}
-\omega_P^2z_{i+1}=0,
\qquad
z_0=z_P=1/\omega_P^2.
\label{eq:s-harmonic-difference}
\end{equation}
The boundary values supply the nonzero entries of $\vect e$ in the
first and last equations.

To find the general solution, substitute $z_i=r^i$ into
Eq.~~\eqref{eq:s-harmonic-difference}.
This gives
\begin{equation}
r^2-\left(2+\omega^2/\omega_P^2\right)r+1=0.
\label{eq:s-harmonic-characteristic}
\end{equation}
The two roots have product one.
Writing the larger root as
\begin{equation}
r=1+\frac{\omega^2}{2\omega_P^2}
+\frac{\omega}{\omega_P}
\sqrt{1+\frac{\omega^2}{4\omega_P^2}},
\label{eq:s-harmonic-root}
\end{equation}
the other root is $r^{-1}$, and the general solution is
$z_i=Cr^i+Dr^{-i}$.

The boundary values in
Eq.~~\eqref{eq:s-harmonic-difference} require
\begin{equation}
C+D=1/\omega_P^2,
\qquad
Cr^P+Dr^{-P}=1/\omega_P^2.
\end{equation}
Solving these equations gives
\begin{equation}
C=\frac{1}{\omega_P^2(1+r^P)},
\qquad
D=\frac{r^P}{\omega_P^2(1+r^P)},
\end{equation}
and therefore
\begin{equation}
z_i=(\vect H^{-1}\vect e)_i
=
\frac{r^i+r^{P-i}}{\omega_P^2(1+r^P)}.
\label{eq:s-harmonic-inverse-components}
\end{equation}
Finally, the definition of $\vect e$ selects the first and last
components of $\vect z$. Expressing their sum in terms of $\omega$ and
$\omega_P$ gives
\begin{equation}
\vect e^T\vect H^{-1}\vect e
=
\frac{2}{\omega_P^2}
\frac{
\cosh\!\left[(P-2)\operatorname{arsinh}
\left(\omega/(2\omega_P)\right)\right]
}{
\cosh\!\left[P\operatorname{arsinh}
\left(\omega/(2\omega_P)\right)\right]
}.
\end{equation}

For brevity, define
$\kappa=\operatorname{arcosh}[1+(\beta_P\hbar\omega)^2/2]$.
The root in Eq.~~\eqref{eq:s-harmonic-root} is then $r=e^\kappa$.
Using this in Eq.~~\eqref{eq:s-harmonic-inverse-components} gives
\begin{equation}
\vect e^T\vect H^{-1}\vect e
=z_1+z_{P-1}
=
\frac{2}{\omega_P^2}
\frac{\cosh[(P-2)\kappa/2]}{\cosh(P\kappa/2)}.
\label{eq:s-harmonic-inverse-element}
\end{equation}

\subsection{Summary of the \texorpdfstring{finite-$P$}{finite-P} expressions}

Substituting Eq.~~\eqref{eq:s-harmonic-inverse-element}
into the expressions derived above gives
\begin{align*}
f_{\qm,P}(q)
&=-\frac{m}{P}
\left[
\omega^2+2\omega_P^2
-\frac{2\omega_P^2\cosh[(P-2)\kappa/2]}
{\cosh(P\kappa/2)}
\right]q,\\
f_{\qm,P}^{\mathrm{RMS}}
&=\left\{
\frac{m}{\beta P}
\left[
\omega^2+2\omega_P^2
-\frac{2\omega_P^2\cosh[(P-2)\kappa/2]}
{\cosh(P\kappa/2)}
\right]
\right\}^{1/2},\\
\left[\sigma_{\qm,P}^{f}(q)\right]^2
&=\frac{2m\omega_P^2}{\beta P}
\frac{\cosh[(P-2)\kappa/2]}{\cosh(P\kappa/2)},\\
\mathrm{NSR}_{\mathrm{prim},P}(q)
&=\left[
\frac{
2\omega_P^2\cosh[(P-2)\kappa/2]
}{
(\omega^2+2\omega_P^2)\cosh(P\kappa/2)
-2\omega_P^2\cosh[(P-2)\kappa/2]
}
\right]^{1/2}.
\end{align*}

\section{Zero-mean total force on unconstrained replicas}
\label{app:zero-mean}

With the first replica fixed at $\vect q^{(1)}=\vect q$,
the total force on replica $j$ is
$\vect f^{(j)}+\vect f^{\spring,(j)}
=-\nabla_{\vect q^{(j)}}U_P$.
For any unconstrained replica, $j=2,\ldots,P$, its conditional mean is
\begin{align}
\left\langle
\vect f^{(j)}+\vect f^{\spring,(j)}
\right\rangle_{\vect q}
&=
\frac{
\displaystyle\int\prod_{k=2}^{P}\textrm{d}\vect q^{(k)}~~
[-\nabla_{\vect q^{(j)}}U_P]e^{-\beta_P U_P}
}{
\displaystyle\int\prod_{k=2}^{P}\textrm{d}\vect q^{(k)}~~
e^{-\beta_P U_P}
}\nonumber\\
&=
\frac{1}{\beta_P}
\frac{
\displaystyle\int\prod_{k=2}^{P}\textrm{d}\vect q^{(k)}~~
\nabla_{\vect q^{(j)}}e^{-\beta_P U_P}
}{
\displaystyle\int\prod_{k=2}^{P}\textrm{d}\vect q^{(k)}~~
e^{-\beta_P U_P}
}
=0.
\label{eq:s-zero-mean}
\end{align}
The final equality follows by integrating over $\vect q^{(j)}$,
provided the boundary contribution vanishes or is periodic.
\section{Optimal weights for the scaled-grad estimator}
\label{app:scaled-grad-weights}

We start from the estimator defined in
Eq.~~\eqref{eq:sg_estimator}:
\begin{equation*}
\vect f_{\qm,P}^{\mathrm{scaled\text{-}grad}}
=
P^{-1}\sum_{j=1}^{P}w^{(j)}
\left[\vect f^{(j)}+\vect f^{\spring,(j)}\right],
\qquad w^{(1)}=1.
\end{equation*}
The weights are fixed during a conditional average at $\vect q$.
Equation~~\eqref{eq:s-zero-mean} shows that the mean total force
on each replica $j=2,\ldots,P$ is zero.
Therefore
\begin{equation}
\left\langle
\vect f_{\qm,P}^{\mathrm{scaled\text{-}grad}}
\right\rangle_{\vect q}
=
P^{-1}
\left\langle
\vect f^{(1)}+\vect f^{\spring,(1)}
\right\rangle_{\vect q}
=
\vect f_{\qm,P}(\vect q).
\label{eq:s-scaled-grad-mean}
\end{equation}
Changing the remaining weights changes the variance of the estimator
without changing its mean.

For any set $\mathcal B$ of trajectory blocks, the sample variance of the block-averaged estimator in Eq.~~\eqref{eq:sg_estimator}, averaged over the $3N$ Cartesian force components, is
\begin{equation}
\operatorname{Var}_{\mathcal B}
\left[\vect f_{\qm,P}^{\mathrm{scaled\text{-}grad}}\right]
=\frac{1}{3N(|\mathcal B|-1)}
\sum_{r\in\mathcal B}
\left|
P^{-1}\sum_{j=1}^{P}w^{(j)}
\left(\vect F_r^{(j)}-\overline{\vect F}_{\mathcal B}^{(j)}\right)
\right|^2.
\label{eq:s-block-variance}
\end{equation}
Equation~~\eqref{eq:s-block-variance} is applied to the validation blocks to select $\lambda$.
To derive the fitted weights, we take $\mathcal B$ to be the fitting blocks in both Eqs.~~\eqref{eq:block-covariances} and \eqref{eq:s-block-variance}.

Expanding Eq.~~\eqref{eq:s-block-variance} gives
\begin{align}
P^2\operatorname{Var}_{\mathcal B}
\left[\vect f_{\qm,P}^{\mathrm{scaled\text{-}grad}}\right]
&=
\operatorname{Var}_{\mathcal B}
\left[\vect F_r^{(1)}\right]
+2\sum_{i=2}^{P}b_iw^{(i)}\nonumber\\
&\quad+
\sum_{i,j=2}^{P}C_{ij}w^{(i)}w^{(j)}.
\label{eq:s-pop-variance}
\end{align}
The first term does not depend on the weights.
Because $\mathbf C$ is symmetric, differentiating the remaining
terms with respect to each free weight gives
\begin{equation}
\frac{\partial}{\partial w^{(i)}}
\left[
2\sum_{j=2}^{P}b_jw^{(j)}
+\sum_{j,k=2}^{P}C_{jk}w^{(j)}w^{(k)}
\right]
=
2b_i+2\sum_{j=2}^{P}C_{ij}w^{(j)},
\qquad i=2,\ldots,P.
\label{eq:s-pop-derivative}
\end{equation}
Setting these derivatives to zero gives the unregularized
minimum-variance condition
\begin{equation}
\sum_{j=2}^{P}C_{ij}w^{(j)}=-b_i,
\qquad i=2,\ldots,P.
\label{eq:s-pop-optimality}
\end{equation}

To fit the weights from finite data, we add the
regularization term used for Eq.~~\eqref{eq:ridge-weights}.
After removing the weight-independent term and the common factor
$P^{-2}$, the quantity minimized is
\begin{equation}
2\sum_{i=2}^{P}b_iw^{(i)}
+\sum_{i,j=2}^{P}C_{ij}w^{(i)}w^{(j)}
+\lambda\sum_{i=2}^{P}[w^{(i)}]^2.
\label{eq:s-ridge-objective}
\end{equation}
Differentiating with respect to $w^{(i)}$ and setting the result
to zero gives
\begin{equation}
\sum_{j=2}^{P}(C_{ij}+\lambda\delta_{ij})w^{(j)}
=-b_i,
\qquad i=2,\ldots,P.
\label{eq:s-ridge-component}
\end{equation}
The vector $\mathbf w(\lambda)$ in Eq.~~\eqref{eq:ridge-weights}
contains precisely these $P-1$ free weights; $w^{(1)}=1$ is fixed
and is not part of the matrix inversion.
Thus Eq.~~\eqref{eq:s-ridge-component} gives
\begin{equation*}
\mathbf w(\lambda)
=-(\mathbf C+\lambda\mathbf I)^{-1}\mathbf b.
\end{equation*}

\section{Zero variance for a harmonic oscillator}
\label{app:harmonic}

We use the harmonic definitions in Appendix~\ref{app:primitive-noise} and write $\delta\vect x=\vect x-\langle\vect x\rangle_q$.
For the free-weight vector $\mathbf w=(w^{(2)},\ldots,w^{(P)})^T$, the fluctuating part of the scaled-grad estimator is
\begin{equation}
 \delta f_{\qm,P}^{\mathrm{scaled\text{-}grad}}
 =\frac{m}{P}(\omega_P^2\vect e-\vect H\mathbf w)^T\delta\vect x.
 \label{eq:s-harmonic-fluctuation}
\end{equation}
The shifted quadratic potential in Appendix~\ref{app:primitive-noise} gives
$\langle\delta\vect x\delta\vect x^T\rangle_q=(\beta_P m)^{-1}\vect H^{-1}$.
Thus the exact conditional variance is
\begin{equation}
 V_{\mathrm{scaled\text{-}grad}}=
 \frac{m}{\beta P}(\omega_P^2\vect e-\vect H\mathbf w)^T
 \vect H^{-1}(\omega_P^2\vect e-\vect H\mathbf w).
 \label{eq:s-harmonic-scaled-var}
\end{equation}
The positive-definite quadratic form in Eq.~\eqref{eq:s-harmonic-scaled-var} vanishes if and only if
\begin{equation}
 \mathbf w=\omega_P^2\vect H^{-1}\vect e.
\end{equation}
Evaluating this inverse gives Eq.~\eqref{eq:ho_optimal_weights}, including $w_{\mathrm{opt}}^{(1)}=1$.
With these weights, the estimator equals $f_{\qm,P}(q)$ in Eq.~\eqref{eq:s-harmonic-effective} for every conditional sample.

\end{document}